\documentclass[osajnl,preprint,showpacs,superscriptaddress,12pt]{revtex4-1} 
\usepackage{amsmath,amssymb,graphicx}
\usepackage{epstopdf} 
\begin{document}

\title{A closed-form solution to predict short-range surface plasmons in thin films and its application to hole arrays.}

\author{Maria Lorente-Crespo}\email{Corresponding author: ml343@hw.ac.uk}
\author{Carolina Mateo-Segura}

\affiliation{Institute of Sensors, Signals and Systems, Heriot-Watt University, EH14 4AS, Edinburgh, UK}

\begin{abstract}A theoretical study of surface plasmon polaritons (SPPs) in ultrathin lossy metal films is presented. The dispersion relation of such films is well known and can be solved numerically to obtain a combination of long-range (LR-) and short-range (SR-) eigenmodes. In this contribution, a simple solution for the SR-SPPs is derived. An approximation for the LR- eigenmodes can be found elsewhere \cite{Yang1991}.  To validate the approximation, a two dimensional (2D) periodic array of small holes is studied subsequently. The spectral response of the array is obtained by full-wave simulations and the results compared with those calculated analytically, showing an excellent agreement for small holes.
\end{abstract}


\maketitle 

When an electromagnetic wave impinges on a metallo-dielectric interface, collective charge oscillations may take place giving rise to bounded modes which propagate along the interface and are confined to its vicinity: SPPs. When two semi-infinite media are considered, the dispersion relation of the SPPs is well known \cite{Raether1988}. The latter may be drammatically different when dealing with finite metal films instead \cite{Economou1969}. In this geometry, SPPs can propagate along both interfaces. For thick films, the modes are degenerate and well described by the dispersion relation of a single flat interface \cite{Raether1988}. Per contra, when the SPPs are separated by a distance smaller than the attenuation length, they can couple across it. As a result, two hybridized modes with thickness-dependent dispersion relation and opposite field distributions appear: the symmetric mode, namely the LR-SPP; and the antysimmetric mode, the SR-SPP. 

In this study we focus in the thin-film case. In this scenario, the aforementioned dispersion relation shows a complex behaviour and cannot be solved analytically. However, if the dielectrics at both sides of the film are equal, we can deal with the LR- and SR- modes separetely. An approximate solution for the LR-SPP dispersion relation can be found in \cite{Yang1991}. Here, we follow a similar approach to derive a closed-form expression of the dispersion relation of the SR-SPPs and apply it to predict the resonance frequencies of a 2D hole array.  Recently, similar arrays have been suggested as absorbers \cite{Bai2010} and polarizers \cite{Braun2009}, based on SR-SPPs resonances. Having a closed solution for calculating the frequencies at which the incident light couples to the SR-SPPs, could simplify and speed-up the design of these and other applications \cite{Xiao2011,Alaverdyan2007,Spevak2009,Zeng2013}.

\begin{figure}[htb]
\centerline{\includegraphics[]{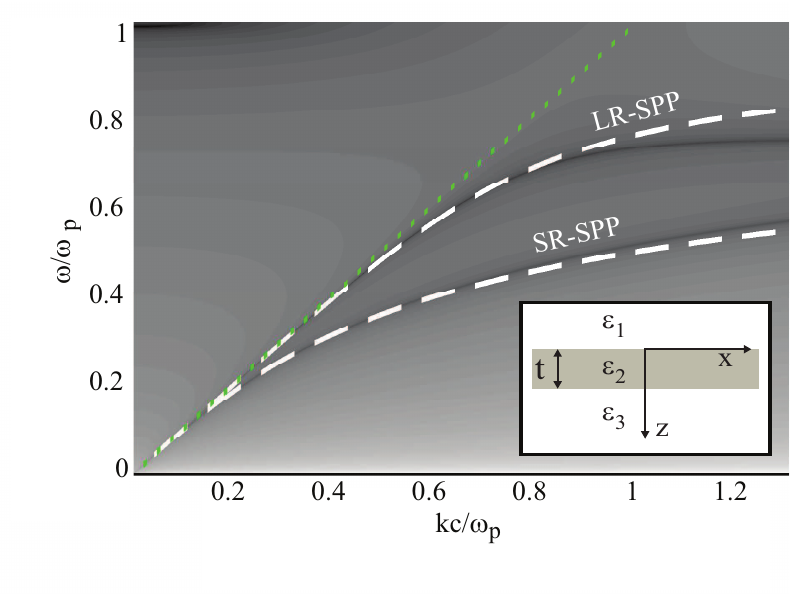}}
\caption{(Color online) Dispersion relation of a 20 nm thick silver film. The light line is shown as a green dotted line. The approximated solutions for the SR- and LR- SPPs are indicated by the white dashed lines. Inset: Metallic film with thickness $t$ and permittivity $\epsilon_2$ sandwiched between two semi-infinite dielectric media with permittivities $\epsilon_1$ and $\epsilon_2$.}
\label{fig1}
\end{figure}

The geometry under analysis is depicted in the inset of Fig. \ref{fig1}. It consists of a metallic slab with thickness $t$ and complex dielectric constant $\epsilon_2=-\epsilon_r-i\epsilon_i$ sandwiched between two semi-infinite  dielectric media with real permittivities $\epsilon_1$ and $\epsilon_3$. Without loss of generality, we let the layers be parallel to the $x$ axis resulting in a one dimensional problem. Since surface plasmon polaritons are transverse magnetic (TM) in nature \cite{Raether1988}, they are better described by their in-plane magnetic field component $\bold{H}_y$:
\begin{equation}
\bold{H}_y=H_0f(z)\exp[i(\omega t-k_x x)]\hat{y},
\end{equation}

where $k_x=k_r-ik_i$ is the in-plane complex propagation constant, $H_0$ is a normalization constant and the propagation is assumed to be in $x$. The term $f(z)$ describes the $z$ dependance of the magnetic field so that it exponentially decays with increasing distance to the interfaces. From the magnetic field, the non-zero electric field components read:
\begin{equation}
\bold{E}_x=\frac{i}{\omega \epsilon_0\epsilon}\frac{\partial H_y}{\partial z}\hat{x},
\end{equation}
\begin{equation}
\bold{E}_z=-\frac{k_x}{\omega\epsilon_0\epsilon}H_y\hat{z}.
\end{equation}

The tangencial components of the magnetic field must be continuous at the interfaces. Therefore, $f(z)$ can be written as:
\begin{widetext}
\begin{equation}
f(z)=
\left\{
    \begin{array}{lr}
     \exp[k_{z_1}z], &z< 0 \\
     \cosh{(k_{z_2}z)}+\frac{k_{z_1}\epsilon_m}{k_{z_2}\epsilon_1}\sinh{(k_{z_2}z)}, &0>z>t \\ 
    \left[\cosh{(k_{z_2}t)}+\frac{k_{z_1}\epsilon_m}{k_{z_2}\epsilon_1}\sin{(k_{z_2}t)}
\right]\exp[-k_{z_3}(z-t)], &z>t
    \end{array}
\right.
\end{equation}
\end{widetext}
where $k_{zi}$ with $i=1,2,3$ are the wavenumbers in the different media and  fulfil the wave equation
\begin{equation}
k_{z_i}^2=k_x^2-\epsilon_i k_0^2.
\end{equation}

Forcing continuity of the tangential electric component, the following dispersion relation is obtained\cite{Burke1986}
\begin{equation}
\tanh{(k_{z_2}t)}(\epsilon_1\epsilon_3 k_{z_2}^2+\epsilon_2^2 k_{z_1} k_{z_3})=-k_{z_2}\epsilon_2(\epsilon_1 k_{z_3}+\epsilon_3 k_{z_1}),
\label{disp}
\end{equation}

which, assuming that the modes are bound in all media, i.e. $\Re[{k_{z_i}}]>0$, in the thick film limit simplifies to that of the single interface
\begin{equation}
k_{SPP}=k_0\sqrt{\frac{\epsilon_2\epsilon_j}{\epsilon_2+\epsilon_j}}, \hspace{20pt} j=1,3.
\label{kspp}
\end{equation}

For a symmetric system, $\epsilon_1=\epsilon_3$, Eq.~\eqref{disp} can be separated in two. One term for symmetric $\bold{H}_y$ and another one for antisymmetric $\bold{H}_y$, which correspond to the LR- and SR-SPP respectively:
\begin{equation}
\tanh{\left(\frac{k_{z_2}t}{2}\right)=\frac{-\epsilon_2 k_{z_1}}{\epsilon_1k_{z_2}}},
\label{long_range}
\end{equation}
\begin{equation}
\tanh{\left(\frac{k_{z_2}t}{2}\right)=\frac{-\epsilon_1 k_{z_2}}{\epsilon_2 k_{z_1}}}.
\label{short_range}
\end{equation}

An approximate solution for the LR-SPP dispersion relation in the thin film limit was already derived in \cite{Yang1991}. Here, we follow a similar approach to find the solution for the SR-SPP case. If the thickness is small enough so that $|k_{z_2}t/2|\ll1$, Eq. \eqref{short_range} can be simplified to
\begin{equation}
\frac{t}{2}\cong\frac{-\epsilon_1}{\epsilon_2 k_{z_1}}.
\label{SR_simplified}
\end{equation}

As $t$ is small, $k_r\to k_0\sqrt{\epsilon_1}$ and $k_i<<1$, we can define $\delta=k_r-k_0 \sqrt{\epsilon_1}$ so that
\begin{equation}
k_{z_1}^2=2k_0\sqrt{\epsilon_1}(\delta-ik_i).
\label{kz1_aprox}
\end{equation}

Substituting Eq.~\eqref{kz1_aprox} in Eq.~\eqref{SR_simplified}, one obtains
\begin{equation}
2k_0\sqrt{\epsilon_1}(k_r-k_0\sqrt{\epsilon_1}-ik_i)-\left[\frac{2\epsilon_1}{(\epsilon_r+i\epsilon_i)d}\right]^2\cong0,
\label{SR_simp_simp}
\end{equation}
whose real and imaginary parts can be treated separately to obtain
\begin{equation}
k_r=\left[\frac{2\epsilon_1}{d(\epsilon_r^2+\epsilon_i^2)} \right]^2 \frac{\epsilon_r^2-\epsilon_i^2}{2k_0\sqrt{\epsilon_1}}+k_0\sqrt{\epsilon_1}
\label{real_SR}
\end{equation}
\begin{equation}
k_i=\frac{\epsilon_r\epsilon_i}{k_0\sqrt{\epsilon_1}}\left[\frac{2\epsilon_1}{d(\epsilon_r^2+\epsilon_i^2)} \right]^2
\label{imag_SR}
\end{equation}


\begin{figure}[tb]
\centerline{\includegraphics[]{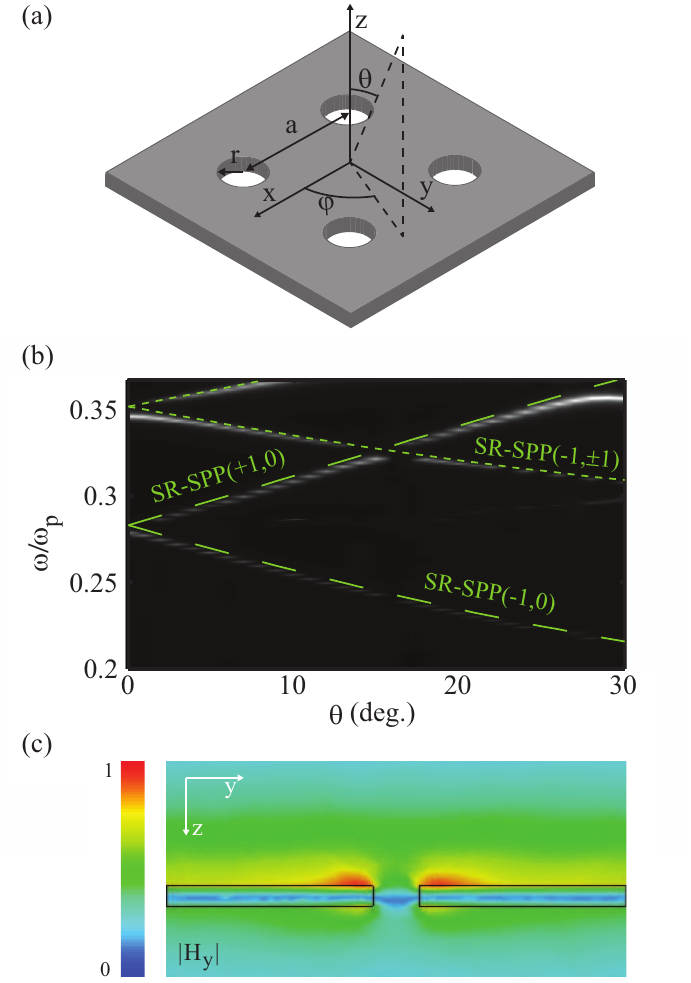}}
\caption{(Color online) (a) Proposed geometry. The plane-wave impinges from $-\hat{z}$. (b) Simulated angular absorption spectra at the $\varphi=0^o$ plane when. Lighter colors correspond to higher values of absorption. The superimposed green lines indicate the theoretical resonance frequencies of the (n,m) order SR-SPPs. (c) Normalized $|H_y|$ distribution at $xz$ plane for $\omega/\omega_p=0.28$ and normal incidence. (b)-(c) $t=20$ nm, \mbox{$a=400$ nm} and $r=20$ nm}
\label{fig2}
\end{figure}

Fig.~\ref{fig1} shows the dispersion relation of a $t=20$ nm silver film surrounded by vacuum ($\epsilon_1=\epsilon_3=1$)  in the real \mbox{$k$-$\omega$} plane. Silver's complex dielectric permittivity is described using a Drude's model
\begin{equation}
\epsilon_r=1-\frac{\omega_p^2}{\omega^2+\omega_c^2},
\end{equation}
\begin{equation}
\epsilon_i=\frac{\omega_c}{\omega}\frac{\omega_p^2}{\omega^2+\omega_c^2},
\end{equation}
 with plasma frequency \mbox{$\omega_p=1.37\times10^{16}$ rad/s} and damping frequency \mbox{$\omega_c=2.5\times10^{14}$ rad/s}. For the sake of comparison, the approximate solution of the dispersion relation of the LR-SPP calculated as in \cite{Yang1991} together with that of the SR-SPP according to Eqs.~\eqref{real_SR} and \eqref{imag_SR} appear superimposed to that obtained by numerically solving Eq.~\eqref{disp}. In the short-k range, the match is excellent. Notice how the dispersion curves lie at the right of the light-line. It is widely accepted that SPPs cannot be directly excited by light impinging from the dielectric because of their larger propagation constant \cite{Raether1988}.  In order to provide the additional momentum necessary to excite the SPPs, in Fig.~\ref{fig2}(a) we include a square grating of subwavelength circular apertures with radius $r$ and periodicity $a$, so that the in-plane component is $\lvert k_{\lvert\lvert}\rvert=\lvert (k_x\pm nG)\hat{x}+(k_y\pm mG)\hat{y}\rvert$, where $G=2\pi/a$ is the reciprocal lattice vector. The approximation given by Eqs. \eqref{real_SR} and \eqref{imag_SR} can be used to analitically preddict the frequencies at which an incoming TM polarized plane wave will couple to the SR-SPPs. 

Fig.~\ref{fig2}(b) shows the simulated angular absorption spectra for a $a=400$ nm grating of $r=35$ nm holes perforated in a $t=20$ nm film when a plane wave is impinging from the $\varphi=0^o$ direction. Absorption is calculated as $1-T-R$, where $T$ and $R$ state for zero-order transmission and reflection, respectively. The cut-off frequency of the holes can be calculated as in \cite{Pfeiffer1974}, by simply exchanging the roles of the dielectric and the metal. For $r=20$ nm holes, $\omega_{c}/\omega_p=1.62$, which is far above the upper frequency limit considered here. Therefore, the holes do not support any propagating modes. Thus, it is reasonable to considered them as weak scatterers. In order to validate the approximation, the SR-SPPs resonances calculated using Eqs.~\eqref{real_SR} and \eqref{imag_SR} are superimposed to the simulations, proving that the derived solution accurately predicts the absorption peaks not only for normal but also for oblique illumination. For off-normal illumination, the SR-SPPs hybridize into even and odd modes which results in the absorption peaks splitting in two. To verify that the excited SP is indeed the SR-SPP, Fig.~\ref{fig2}(c) shows the magnitude of the paralell component of the magntic field, $H_y$, normalized to its maximum at $\omega/\omega_p=0.28$ and normal incidence. As expected for the SR-SPPs, $H_y$ exhibits a zero inside the metal film.

\begin{figure}[tb]
\centerline{\includegraphics[]{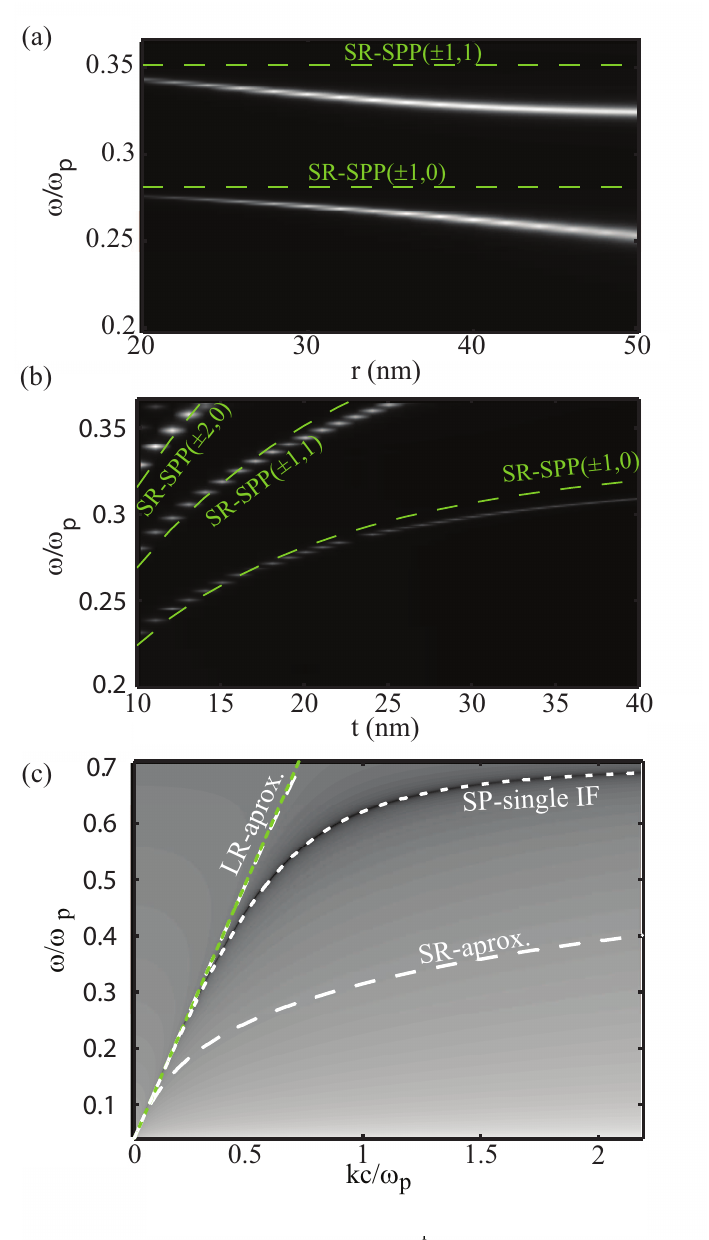}}
\caption{(Color online) Angle resolved absorption at normal incidence for a $a=400$ nm grating (a) vs. hole radius when $t=20$ nm; (b) vs. thickness of the film when $r=20$ nm. The predicted frequencies are shown as green dashed lines. (c) Dispersion relation of a 300 nm thick silver film. The light line is shown as a green dotted line. The approximated solutions for the SR- and LR- SPPs are indicated by the white dashed lines, while that of the single interface corresponds to the white dotted line.}
\label{fig3}
\end{figure}

The effect of increasing the size of the apertures is investigated in Fig.~\ref{fig3}(a). Up to this point any effects due to the holes, appart from providing a mechanism for the excitation of SPPs, have been neglected. For larger holes, the simulations increasingly deviate from the theory as a consequence of the empty lattice approximation limited validity. The red-shift of the resonances with respect to the predictions has been extensively studied in the literature \cite{Lezec2004,Pacifici2008,Liu2008} and is out of the scope of this work.

The thickness of the metal film plays a key role in the dispersion relation of thin films and thus, in the excitation of the SR-SPPs. Moreover, the approximation derived in this work is thickness dependent and remains valid as long as $|k_{z_2}t/2|\ll1$. Therefore, the estimated resonances are more accurate for thinner films as confirmed by Fig.~\ref{fig3}(b). Above a critical thickness, the SR and LR-SPPs are uncoupled and converge to the SP mode of the single interface, Eq.~\eqref{kspp}, which is not well described by Eqs.~\eqref{real_SR} and \eqref{imag_SR}. To ilustrate this, the dispersion relation given by Eq.~\eqref{disp} is solved for a $t=300$ nm film. The predicted normalized propagation constants of the LR- and SR-SPPs according to \cite{Yang1991} and Eqs.~\eqref{real_SR} and \eqref{imag_SR}, respectively, appear superimposed in Fig.~\ref{fig3}(c), together with that of the single interface. Notice that the LR- and SR- SPP are degenerate in this case and Eq.~\eqref{kspp} is more accurate for describing the overall behaviour.


In conclusion, we derived a simple expression for calculating the dispersion relation of SR-SPPs of a thin lossy metal film which can be used to predict the coupling of light to such SPs and hence facilitate the design of SR-SPPs based applications. To validate the approximation, we compared the estimated resonance frequencies to the ones obtained when solving the transcendental equation for finite metal slabs, finding an excellent agreement in the short-k region. To further explore the accuracy of this approach we also considered a lattice of subwavelength holes as a mechanism to excite the SR-SPPs. Within the empty lattice approximation limitations, the estimated and simulated resonance frequencies are in very good agreement for normal and oblique incident light. Our solution remains valid as long as $|k_{z_2}t/2|\ll1$. Above a critical thickness, the SR- and LR- modes become degenerate and the dispersion curves converge to that of a single smooth interface as suggested by the simulations.


%

%
%

\end{document}